\documentstyle[aps,prd,eqsecnum,epsf]{revtex}

\begin{document}

\newcommand{\gsim}{\mbox{\raisebox{-1.ex}{$\stackrel
     {\textstyle>}{\textstyle\sim}$}}}
\newcommand{\lsim}{\mbox{\raisebox{-1.ex}{$\stackrel
     {\textstyle<}{\textstyle \sim}$}}}
\newcommand{\square}{\kern1pt\vbox{\hrule height
1.2pt\hbox{\vrule width 1.2pt\hskip 3pt
   \vbox{\vskip 6pt}\hskip 3pt\vrule width 0.6pt}\hrule
height 0.6pt}\kern1pt}

\newcommand{\singlefig}[2]{
\begin{center}
\begin{minipage}{#1}
\epsfxsize=#1
\epsffile{#2}
\end{minipage}
\end{center}}
%
\newenvironment{figcaption}[2]{
 \vspace{0.3cm}
 \refstepcounter{figure}
 \label{#1}
 \begin{center}
 \begin{minipage}{#2}
 \begingroup \small FIG. \thefigure: }{
 \endgroup
 \end{minipage}
 \end{center}}
%


\draft
\title{{\bf New constraints on multi-field inflation 
with nonminimal coupling\\}}
\author{Shinji Tsujikawa\thanks{electronic
address: shinji@gravity.phys.waseda.ac.jp}
and Hiroki Yajima\thanks{electronic
address: yajima@gravity.phys.waseda.ac.jp}}
\address{Department of Physics, Waseda University,
Shinjuku-ku, Tokyo 169-8555, Japan\\[.3em]}
\date{\today}
\maketitle
\begin{abstract}
We study the dynamics and perturbations
during inflation and reheating in a multi-field model 
where a second scalar field $\chi$ is nonminimally coupled to 
the scalar curvature $(\frac12 \xi R\chi^2$). When $\xi$ is positive, 
the usual inflationary prediction for large-scale anisotropies is hardly 
altered while the $\chi$  fluctuation in sub-Hubble
modes can be amplified during preheating for large $\xi$.
For negative values of $\xi$, however, 
long-wave modes of the $\chi$ fluctuation exhibit exponential 
increase during inflation, leading to the strong enhancement of 
super-Hubble metric perturbations even when $|\xi|$ is 
less than unity. This is because the effective $\chi$ mass becomes
negative during inflation. We constrain the strength of $\xi$  and 
the initial $\chi$ by the amplitude of produced 
density perturbations. One way to avoid nonadiabatic growth of 
super-Hubble curvature perturbations is to stabilize the $\chi$ mass 
through a coupling to the inflaton. 
Preheating may thus be necessary in these models to protect 
the stability of the inflationary phase.   
\end{abstract}

\pacs{PACS 98.80.Cq}

\baselineskip = 20pt

\section{Introduction}                           %

The idea of inflation is remarkable in the sense that it can not only 
solve the horizon and flatness problems of the standard big bang cosmology,
but provides seeds of density perturbations relevant for the 
large scale structure\cite{Kolb}. The perturbations give an imprint on the 
Cosmic Microwave Background (CMB) anisotropies, 
whose temperature fluctuations can be analyzed by present observations.
The inflationary paradigm typically predicts the nearly scale-invariant 
primordial power spectrum\cite{den1,den2}, which is consistent with 
observations of Cosmic Background Explorer (COBE) satellite.
Since the accuracy of measurements is expected to be improved in 
future observations, it is very important to fully understand
the primordial power spectrum predicted by the inflationary paradigm.

Generally, it is assumed that only one scalar field called 
{\it inflaton} determines the dynamics of inflation, which leads to 
the exponential expansion of the universe when inflaton slowly 
evolves along the sufficiently flat potential.
In the single-field model, density perturbations are typically ``frozen''
when a physical scale crosses the Hubble radius during inflation.
This makes it possible to evaluate the power spectrum at the end of 
inflation by equating it at the first horizon crossing.
In preheating era after inflation, the fluctuation of inflaton 
can be enhanced by parametric resonance\cite{pre1,pre2}, which 
may stimulate the growth of metric perturbations. 
In the single-field case, however, the super-Hubble curvature perturbation 
is typically conserved during preheating\cite{mpre1}, 
while sub-Hubble modes can be amplified in some models 
of inflation\cite{mpre2} including the nonminimally coupled
inflationary model\cite{nonminiinflaton}. 
As long as the system is a single-field model,
and the stress-energy is conserved,  nonadiabatic growth of 
the large-scale curvature perturbation can not be expected 
during inflation and reheating including 
generalized Einstein theories\cite{Hwang}.

Multi-field inflationary scenarios have received much 
attention for the generality of inflation and preheating.
In fact, density perturbations in multi-field models 
were analytically derived by several authors in the scheme of the 
slow-roll approximation\cite{multi}.
In the presence of more than two scalar fields, large-scale 
curvature perturbations are not necessarily conserved due to 
the existence of isocurvature perturbations.
In the context of scalar-tensor gravity theories, several 
authors\cite{multi2,multi3} studied density perturbations in the two-field system
where there exists a Brans-Dicke or dilaton field in addition to inflaton.  
In particular, Garc\'{\i}a-Bellido and Wands \cite{multi3} constrained 
parameters of the gravity theories by comparing the predicted spectral index 
with observational datas. 
In addition to this, since the higher-dimensional generalized Kaluza-Klein 
theories\cite{KK} also give rise to a dilaton field by reducing the effective 
four-dimensional theories,  it is worth investigating
to predict the primordial power spectrum in the presence of inflaton in 
generalized Einstein theories from a cosmological point of view.
In this respect, Berkin and Maeda\cite{BM} studied the 
new and chaotic inflationary models with a dilaton potential 
$U(\sigma)=0$, and constrained the parameters of models by produced
density perturbations. Multiple scalar fields
also play important roles in the assisted inflation with exponential 
potentials\cite{assisted}. This scenario was recently extended to the assisted 
chaotic inflation induced by higher-dimensional theories\cite{assisted2}, and 
density perturbations were calculated in Ref.\cite{assisted3}.

In preheating era, scalar fields coupled to inflaton can be 
resonantly amplified, which is typically more efficient 
than in the single-field case. It was also pointed out that there is also
an interesting possibility that super-Hubble metric perturbations
will be excited due to the growth of field 
perturbations\cite{mpre3,mpre4,mpre5}.
Since growth of metric perturbations can be expected as long as
scalar fields are not severely damped in the inflationary 
period~\cite{suppress,suppress2} and are enhanced during preheating, 
it is important to take care the dynamics of scalar fields during 
inflation. When the effective mass of scalar fields is heavy relative to 
the Hubble parameter $H$, long wave modes of field fluctuations exhibit 
exponential decrease during inflation~\cite{suppress}.
In contrast, ``light'' fields such as inflaton whose masses are 
smaller than $H$ are hardly affected by the 
inflationary suppression~\cite{mpre4}, 
and can lead to the enhancement of super-Hubble metric perturbations
in preheating era if they are amplified by parametric resonance\cite{mpre5}. 
In this respect, one of the present authors recently 
investigated the evolution of field and metric perturbations
in the presence of a dilaton field with quadratic
inflaton potential\cite{shinji}, and found that the curvature 
perturbation in cosmological relevant modes remains almost 
constant in this model (Candelas Weinberg model, see Ref.~\cite{CW}), 
including the backreaction effect of created particles.

From the viewpoint of quantum field theories
in curved spacetime, nonminimal couplings naturally arise,
with their own nontrivial renormalization group flows. 
The ultra-violet fixed point of these flows are often divergent, 
implying that nonminimal couplings may be important 
at high energies\cite{BOS}.
In the single-field case with a nonminimally coupled inflaton field, 
Futamase and Maeda\cite{FM} studied the dynamics of chaotic 
inflation, and found that the nonminimal coupling is constrained as 
$|\xi|~\lsim~10^{-3}$ in the quadratic potential,
by the requirement of sufficient amount of inflation.
On the other hand, such a constraint is absent in the self-coupling 
potential for negative $\xi$, and as a bonus, the fine tuning problem of 
the self-coupling $\lambda$ in the minimally coupled case 
can be relaxed by large negative values of $\xi$\cite{SBB,FU}.
Several authors evaluated  scalar and tensor perturbations 
generated during inflation\cite{MS,spectral,KF} and 
preheating\cite{nonminiinflaton} in this model. 
Since the system is reduced to the single-field model with some modified
inflaton potential by a conformal transformation, 
the super-Hubble curvature perturbation remains almost constant, while  
metric preheating is found to be vital in sub-Hubble scales\cite{nonminiinflaton}.

In the multi-field model with inflaton and a nonminimally 
coupled scalar field $\chi$, it was found that $\chi$ particles 
can be efficiently produced {\it during  inflation} when $\xi$ is 
negative in the unperturbed Friedmann-Robertson-Walker 
background\cite{SH}. 
The dynamics of scalar fields strongly depends
on the coupling $\xi$. In fact, although the exponential suppression
of super-Hubble $\chi$ modes will take place
for positive $\xi$ due to large effective mass relative to the Hubble rate,
they can grow exponentially by negative instability for $\xi<0$. 
Then it is expected that negative 
nonminimal coupling may lead to the enhancement of 
super-Hubble metric perturbations during inflation.
In addition to this, it is of interest how metric preheating 
proceeds in large scales, since the $\chi$ fluctuation can also 
be amplified by parametric resonance when $|\xi|$ is greater than 
of order unity\cite{nonminimalpre2,nonminimalpre3}.
In this paper, motivated by above considerations, we will 
make precise analysis about the evolution of field and metric 
perturbations during inflation and preheating
in the presence of a nonminimally coupled scalar field $\chi$.
We believe that our study will be important in the sense that 
we can constrain  the strength of nonminimal coupling
by the COBE normalization.
In the case where the power spectrum exceeds the observational 
upper bound by negative nonminimal coupling, we will give 
one escape route from nonadiabatic growth of super-Hubble 
metric perturbations.

 \section{The model and basic equations}
We investigate a model where a massless scalar field $\chi$ is
nonminimally coupled with the scalar curvature $R$
in the presence of an inflaton field $\phi$:
\begin{eqnarray}
 {\cal L}= \sqrt{-g} \left[ \frac{1}{2\kappa^2}R
   -\frac{1}{2}(\nabla \phi)^2
   -V(\phi) 
   -\frac{1}{2}(\nabla \chi)^2
   -\frac12 \xi R \chi^2
    \right], 
\label{B1}
\end{eqnarray}
where $G \equiv \kappa^{2}/8\pi =m_{\rm pl}^{-2}$ is a
gravitational coupling constant, and $\xi$ is a nonminimal
coupling. In this paper, we adopt the quadratic potential  
for inflaton,
\begin{eqnarray}
 V(\phi)=\frac12 m^2\phi^2.
\label{B2}
\end{eqnarray}
The variation of the action Eq.~$(\ref{B1})$ yields the following field
equations:
\begin{eqnarray}
 \frac{1-\xi\kappa^2\chi^2}{\kappa^2}G_{\mu\nu} &=&
 2\xi\chi(g_{\mu\nu}\square\chi-\nabla_{\mu}\nabla_{\nu}\chi)
 -g_{\mu\nu}V(\phi)+(\nabla_{\mu}\phi)(\nabla_{\nu}\phi)-
 \frac{1}{2}g_{\mu\nu} (\nabla^{\lambda}\phi)
 (\nabla_{\lambda}\phi) \nonumber \\
&+& (1-2\xi)(\nabla_{\mu}\chi)(\nabla_{\nu}\chi)-
\left(\frac{1}{2}-2\xi \right)
g_{\mu\nu} (\nabla^{\lambda}\chi)(\nabla_{\lambda}\chi),
\label{B4}
\end{eqnarray}
\begin{eqnarray}
\square\phi-V'(\phi)=0,
\label{B5}
\end{eqnarray}
\begin{eqnarray}
\square\chi-\xi\chi R=0,
\label{B6}
\end{eqnarray}
where a prime denotes the derivative with respect to $\phi$.

Let us consider the perturbed metric in the longitudinal gauge
around a Friedmann-Lema$\hat{\i}$tre-Robertson-Walker
(FLRW) background
\begin{eqnarray}
ds^2=-(1+2\Phi)dt^2
+a^2(t)(1-2\Psi)\delta_{ij} dx^i dx^j,
\label{B3}
\end{eqnarray}
with $a(t)$ the scale factor, and $\Phi, \Psi$ are 
gauge-invariant potentials\cite{den2}.
Decomposing  scalar fields into $\varphi_J(t,{\bf
x}) \to \varphi_J (t)+\delta\varphi_J(t,{\bf x})~
(J=1, 2)$, where $\varphi_J (t)$ are homogeneous parts
and $\delta\varphi_J(t,{\bf x})$ are
gauge-invariant fluctuations, we obtain the following
background equations for the Hubble parameter 
$H \equiv \dot{a}/a$ and scalar fields $\varphi_J (t)$:
\begin{eqnarray}
H^2=\frac{\kappa^2}{3(1-\xi\kappa^{2}\chi^{2})}
\left[\frac{1}{2}\dot{\phi}^2+V(\phi)+\frac{1}{2}
\dot{\chi}^2+6\xi H \chi\dot{\chi} \right],
\label{B7}
\end{eqnarray}
\begin{eqnarray}
\dot{H}=-\frac{\kappa^2}{2(1-\xi\kappa^{2}\chi^{2})}
\left[\dot{\phi}^2+(1-2\xi)\dot{\chi}^2-2\xi\chi
(\ddot{\chi}-H\dot{\chi}) \right],
\label{B60}
\end{eqnarray}
\begin{eqnarray}
\ddot{\phi}+3H\dot{\phi}+V'(\phi)=0,
\label{B8}
\end{eqnarray}
\begin{eqnarray}
\ddot{\chi}+3H\dot{\chi}+\xi R\chi =0,
\label{B9}
\end{eqnarray}
where the scalar curvature $R$ is given by 
\begin{eqnarray}
R &=& 6(2H^2+\dot{H}) \nonumber \\
&=& \frac{\kappa^2}{1-\xi\kappa^{2}\chi^{2}}
\left[-\dot{\phi}^2+4V(\phi)-\dot{\chi}^2+18\xi H \chi \dot{\chi}
+6\xi(\dot{\chi}^2+\chi\ddot{\chi}) \right].
\label{B10}
\end{eqnarray}
The Fourier modes of the linearized perturbed
Einstein equations are written as\cite{Hwang}
\begin{eqnarray}
\Psi_k=\Phi_k-\frac{2\xi\kappa^2\chi}
{1-\xi\kappa^{2}\chi^{2}} \delta\chi_k,
\label{B11}
\end{eqnarray}
\begin{eqnarray}
\dot{\Psi}_k+\left(H-\frac{\xi\kappa^2\chi\dot{\chi}}
{1-\xi\kappa^{2}\chi^{2}}\right) \Phi_k=\frac{\kappa^2}
{2(1-\xi\kappa^{2}\chi^{2})} \left[\dot{\phi}\delta\phi_k+
(1-2\xi)\dot{\chi}\delta\chi_k-2\xi\chi(\delta\dot{\chi}_k
-H\delta\chi_k) \right],
\label{B12}
\end{eqnarray}
\begin{eqnarray}
\delta\ddot{\phi}_k+3H\delta\dot{\phi}_k+
\left[ \frac{k^2}{a^2} +V''(\phi) \right] \delta\phi_k=
2(\ddot{\phi}+3H\dot{\phi})\Phi_k+
\dot{\phi}(\dot{\Phi}_k+3\dot{\Psi}_k),
\label{B13}
\end{eqnarray}
\begin{eqnarray}
\delta\ddot{\chi}_k+3H\delta\dot{\chi}_k+
\left( \frac{k^2}{a^2} +\xi R \right)\delta \chi_k=
2(\ddot{\chi}+3H\dot{\chi})\Phi_k
+\dot{\chi}(\dot{\Phi}_k+3\dot{\Psi}_k)
-\xi\chi \delta R_k,
\label{B14}
\end{eqnarray}
with 
\begin{eqnarray}
\delta R_k=-\left[12(2H^2+\dot{H})\Phi_k+
6 H (\dot{\Phi}_k+4\dot{\Psi}_k)+6\ddot{\Psi}_k
-\frac{2k^2}{a^2}\Phi_k+\frac{4k^2}{a^2}\Psi_k
\right].
\label{B15}
\end{eqnarray}
Note that $\Phi_k$ and $\Psi_k$ do not coincide in the 
nonminimally coupled case, due to the
nonvanishing anisotropic stress.
In the absence of the nonminimally coupled $\chi$ field 
(i.e. single-field case), there exists a conserved quantity 
$\zeta_k \equiv -\Psi_k+(\Psi_k+\dot{\Psi}_k/H)H^2/\dot{H}$
for super-Hubble $k$ modes in the linear perturbations\cite{den2}.
During reheating phase, although entropy perturbations
can be produced when $\dot{\phi}$ periodically passes through 
zero, curvature perturbations in super-Hubble scales 
are typically conserved in single-field models\cite{mpre1} 
including the nonminimally coupled inflaton case\cite{nonminiinflaton}.
Even in  generalized Einstein theories including scalar-tensor 
and higher-curvature gravity theories, it was found that the conserved
structure in large scales still holds in the single-field case\cite{Hwang}.
Due to this adiabaticy of the curvature perturbation, we only take care  
the perturbation at the first horizon crossing in order to evaluate the
inflationary power spectrum.

In the multi-field case, however, the curvature perturbation 
on uniform-density hypersurfaces\cite{Hwang},
\begin{eqnarray}
\zeta_k \equiv -\Psi_k+\frac{H^2(1-\xi\kappa^2\chi^2)}
{\dot{H}(1-\xi\kappa^2\chi^2)+2H\xi\kappa^2\chi\dot{\chi}}
\left(\Psi_k+\frac{\dot{\Psi}_k}{H}\right),
\label{C40}
\end{eqnarray}
includes the isocurvature perturbation\cite{multi,multi2,multi3}, 
which can vary nonadiabatically during inflation and reheating.
In fact, in the present model, since 
the homogeneous $\chi$ and the $\delta\chi_k$ fluctuation
in small $k$-modes can be strongly excited for negative $\xi$ 
as is found in Eqs.~$(\ref{B9})$ and $(\ref{B14})$, 
this will stimulate the growth of super-Hubble metric 
perturbations and produce entropy perturbations.
On the other hand, for positive $\xi$,
it is expected that long-wave $\chi$ modes
will exponentially decrease during inflation due to the large effective 
$\chi$ mass relative to the Hubble rate\cite{suppress,suppress2}, 
which may not lead to the nonadiabatic growth of curvature perturbations 
on super-Hubble scales even if field fluctuations will exhibit parametric 
amplifications in reheating phase.
In the next section, we will make a detailed analysis about 
the dynamics of field and metric perturbations during inflation 
and reheating.

 \section{Cosmological perturbations during inflation and reheating}
Let us first review the dynamics of inflation with potential $(\ref{B2})$
in the absence of the nonminimally coupled $\chi$ field.
Neglecting the $\dot{\phi}$ term in Eq.~$(\ref{B7})$ and
the $\ddot{\phi}$ term in Eq.~$(\ref{B8})$, we obtain the following
approximate relation during inflation:
\begin{eqnarray}
H \approx \sqrt{\frac{4\pi}{3}}\frac{m}{m_{\rm pl}}\phi,~~~~
\dot{\phi} \approx -\frac{m^2}{3H}\phi.
\label{C1}
\end{eqnarray}
Combining these relations is to give
\begin{eqnarray}
\phi=\phi(0)-\frac{m_{\rm pl}}{\sqrt{12\pi}}mt,
\label{C2}
\end{eqnarray}
\begin{eqnarray}
a=a(0) \exp \left[ \sqrt{\frac{4\pi}{3}} \frac{m}{m_{\rm pl}}
\left\{ \phi(0)t-\frac{m_{\rm pl}}{\sqrt{48\pi}}mt^2 \right\}
\right],
\label{C3}
\end{eqnarray}
where $\phi(0)$ and $a(0)$ are initial values of inflaton and
the scale factor, respectively. In the initial stage of inflation, 
the scale factor evolves exponentially as
$a \sim a(0) \exp [ \sqrt{4\pi/3}(m/m_{\rm pl})\phi(0) t ]$. 
With the increase of the last term in 
Eq.~$(\ref{C3})$, the expansion rate slows down, which is 
followed by the oscillating stage of inflaton.
In order to solve several cosmological puzzles of the standard
big bang cosmology, the number of 
e-foldings $N \equiv \ln (a/a(0))$ is required to be $N~\gsim~55$,
by which the initial value of inflaton is constrained as 
$\phi(0)~\gsim~3m_{\rm pl}$. 
The inflationary period ends when the slow-roll parameter 
$\epsilon \equiv (V'/V)^2/2\kappa^2$ grows of order unity, which 
corresponds to $\phi \sim 0.3 m_{\rm pl}$.

If nonminimal coupling is taken into account, the dynamics of inflation 
can be changed. In fact, growth of the $\chi$ field affects
the Hubble parameter by Eq.~$(\ref{B7})$, which also alters 
the evolution of inflaton by Eq.~$(\ref{B8})$.
Let us first investigate the evolution of the $\chi$ 
fluctuation approximately.
Neglecting the contribution of metric perturbations in Eq.~$(\ref{B14})$
and introducing a new scalar field $\delta X_k \equiv a \delta\chi_k$ and 
a conformal time $\eta \equiv \int a^{-1}dt$, Eq.~$(\ref{B14})$ 
yields
\begin{eqnarray}
\delta X_k''+\left[ k^2-(1-6\xi)\frac{a''}{a}\right]\delta X_k=0,
\label{C6}
\end{eqnarray}
where a prime denotes the derivative with respect to the conformal time.
This solution can be expressed by the combinations of the Hankel
functions $H_{\nu}^{(J)}$ ($J=1, 2$)\cite{SH}:
\begin{eqnarray}
\delta\chi_k=a^{-1} [c_1 \sqrt{\eta} H_{\nu}^{(2)}(k \eta)
+c_2 \sqrt{\eta} H_{\nu}^{(1)}(k \eta)],
\label{C4}
\end{eqnarray}
where the order $\nu$ of the Hankel functions is given by
\begin{eqnarray}
\nu^2=\frac94 -12\xi.
\label{C5}
\end{eqnarray}
Note that the choice of $c_1=\sqrt{\pi}/2$ and $c_2=0$ corresponds
to the state of the Bunch-Davies vacuum.
The solution of the homogeneous $\chi$ field also looks the
form of Eq.~$(\ref{C4})$ with $k=0$.

The Hankel functions take the following form in the limit of 
$k\eta \to 0$\cite{Kaiser}:
\begin{eqnarray}
H_{\nu}^{(2,1)}(k\eta) \to \pm \frac{i}{\pi} \Gamma(\nu)
\left(\frac{k\eta}{2}\right)^{-\nu}, 
\label{C7}
\end{eqnarray}
where $\Gamma(\nu)$ is the Gamma function with order $\nu$.
Taking notice of the relation $\eta \approx -1/(aH)$ during inflation,
we easily find from Eq.~$(\ref{C4})$ that long-wave
$\delta\chi_k$ modes exponentially increase as $\delta\chi_k \sim
a^{\nu-3/2}$ when $\nu>3/2$.
This case corresponds to the negative $\xi$ by Eq.~$(\ref{C5})$, 
leading to the particle creation during inflation by negative instability  
as noted in Ref.~\cite{SH}.
This efficient particle production for low momentum modes 
is expected to enhance metric perturbations
for wavelengths larger than the Hubble radius,
due to the increase of the rhs of Eq.~$(\ref{B12})$.
This will also stimulate the growth of field perturbations 
as expected by Eqs.~$(\ref{B13})$ and $(\ref{B14})$.
In contrast, for positive $\xi$, long-wave
$\delta\chi_k$ modes decay exponentially in 
de Sitter background. Especially for the case of 
$\xi>3/16$ where $\nu$ takes complex values, $\delta\chi_k$ 
decreases as $\sim a^{-3/2}$.
This makes the $\chi$ term in the rhs of
Eq.~$(\ref{B12})$ unimportant and super-Hubble metric 
perturbations will not be strongly amplified even taking into
account the parametric amplification of $\chi$ and $\delta\chi_k$ modes
during preheating as shown in Ref.~\cite{suppress} in the model of 
$V(\phi,\chi)=\frac12 m^2\phi^2+\frac12 g^2\phi^2\chi^2$.

Before analyzing the dynamics of the system, we should mention
initial conditions of field fluctuations.
For positive frequency $\omega_k^2>0$, we typically choose the 
conformal vacuum state $\delta\varphi_k=1/\sqrt{2\omega_k}$
and $\delta\dot{\varphi}_k=(-i\omega_k -H)\delta\varphi_k$. 
However, in de Sitter background with nonminimal coupling,
the frequency of scalar fields can take negative values.
Hence we adopt the de Sitter invariant vacuum 
state given by\cite{BD80}
\begin{eqnarray}
\delta X_k(\eta_0)=\frac{1}{k^{3/2}}\left[ ik+\frac{a'(\eta_0)}{a(\eta_0)}
\right] \exp(ik\eta_0),~~~\delta X_k'(\eta_0)=
\frac{1}{k^{1/2}}\left[ ik+\frac{a'(\eta_0)}{a(\eta_0)}
-\frac{i}{k}\left(\frac{a'(\eta_0)}{a(\eta_0)}\right)'
 \right]\exp(ik\eta_0).
\label{C8}
\end{eqnarray}
In the case of $k \gg a'(\eta_0)/a(\eta_0)$, this takes the similar
form as the conformal vacuum state, while mode functions 
depend on the choice of the vacuum for small $k$.
However, in the context of the inflationary power spectrum, it is known 
that different choice of initial conditions has little affect
on the results\cite{den2}.

In what follows, we numerically study the evolution of
scalar fields and super-Hubble metric perturbations for positive
and negative values of $\xi$ during inflation and reheating, and 
also investigate the case where the coupling between
$\phi$ and $\chi$~($\frac12 g^2\phi^2\chi^2$) is introduced
at the end of this section.

 \subsection{Case of $\xi>0$}
Let us first consider the minimally coupled case ($\xi=0$)
before analyzing the positive $\xi$ case. 
In this case, the inflationary period proceeds as in the 
single-field case, as long as $\chi$ is initially small
relative to inflaton. 
We plot in Fig.~1 the evolution of 
the metric perturbation $\Psi_k (=\Phi_k)$ and the curvature 
perturbation $\zeta_k$ for a cosmological mode $k=a_0H_0$
where $a_0$ denotes the scale factor about 55 e-foldings
before the end of inflation. 
The initial value of inflaton is chosen as
$\phi(0)=3m_{\rm pl}$, in which case the inflationary period 
continues until $mt \approx 20$, leading to about 57 e-foldings
at the end of inflation. 
The curvature perturbation,
\begin{eqnarray}
\zeta_k = -\Psi_k+\frac{H^2}{\dot{H}}
\left(\Psi_k+\frac{\dot{\Psi}_k}{H}\right),
\label{C41}
\end{eqnarray}
remains almost constant on large-scales as is clearly seen in 
Fig.~1.

When the system enters the reheating stage, $\dot{\phi}$
periodically passes through zero, during which entropy 
perturbations can be produced\cite{mpre3}.
Nevertheless, this process is not strong 
enough to lead to the overall increase of $\zeta_k$ and $\Phi_k$
in super-Hubble modes (see Fig.~1).
Not only scalar field fluctuations 
for low momentum modes are not relevantly amplified, 
but metric preheating is inefficient on large scales for $\xi=0$.

Let us define the power-spectrum of $\zeta_k$ as
\begin{eqnarray}
{\cal P}_{\zeta_k}\equiv 
 \frac{k^3}{2\pi^2}|\zeta_k|^2=\frac{|\tilde{\zeta}_k|^2}{2\pi^2},
\label{C9}
\end{eqnarray}
where $\tilde{\zeta}_k$ is defined by 
$\tilde{\zeta}_k \equiv k^{3/2}\zeta_k$.
Assuming that $\zeta_k$ remains conserved in cosmological 
scales until it reenters the Hubble length in the  
matter-dominant stage, the power spectrum 
at the end of inflation [$={\cal P}_{\zeta_k}(t_e)$] can be 
related with the density perturbation $\delta_H(k)$ 
at the horizon reentry as\cite{den2,multi3}
\begin{eqnarray}
\delta_H^2(k) \approx \frac{4}{25} {\cal P}_{\zeta_k}(t_e).
\label{C25}
\end{eqnarray}
For the inflaton mass $m \sim 10^{-6}m_{\rm pl}$
regulated by CMB observations, numerical calculations give
$\tilde{\zeta}_k \approx 2.4 \times 10^{-4}$ at the end of inflation.
Then we obtain the density perturbation as $\delta_H(k)
\approx 2 \times 10^{-5}$ by the relation $(\ref{C25})$.

\begin{figure}
\begin{center}
\singlefig{12cm}{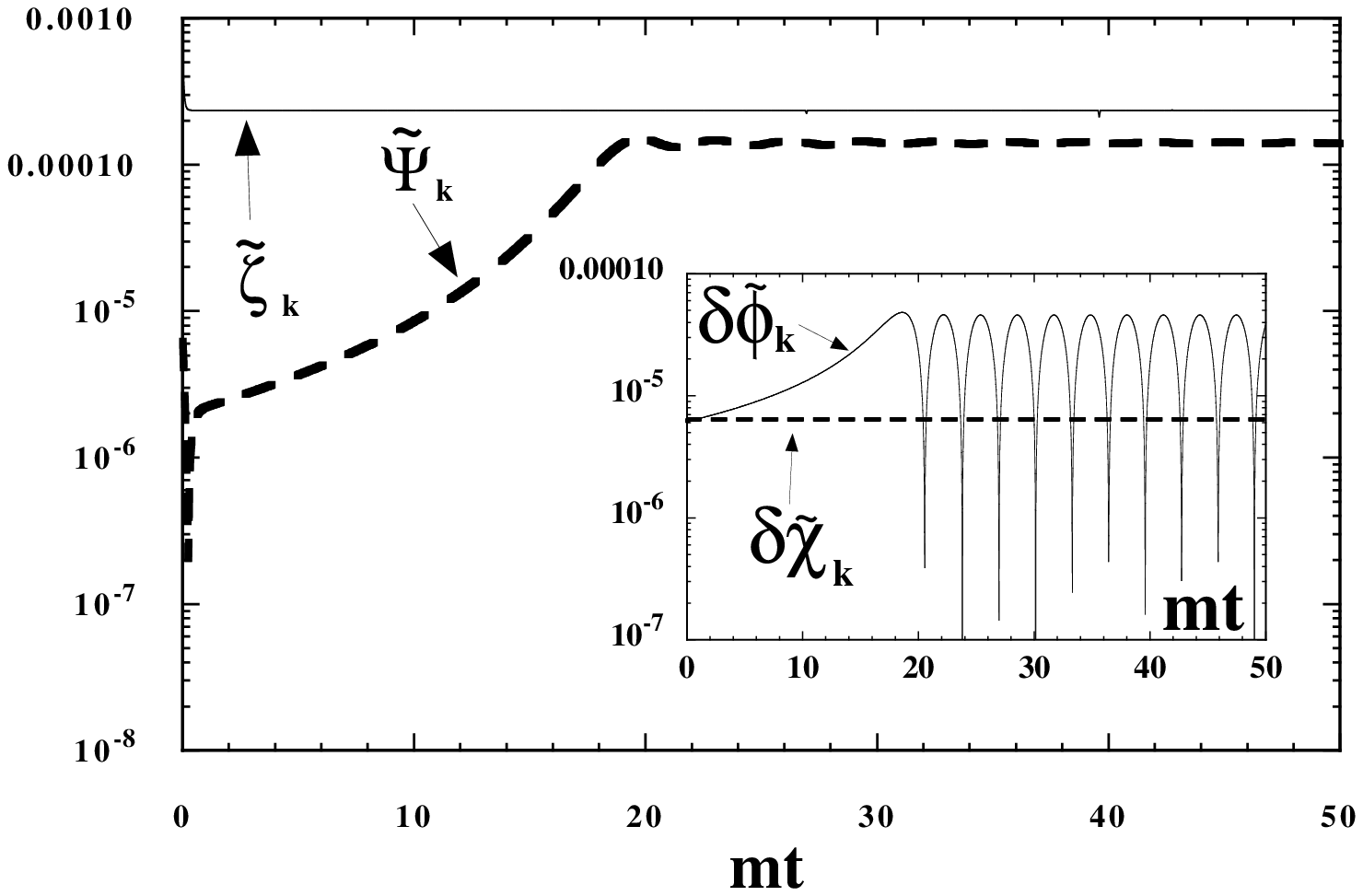}
\begin{figcaption}{Fig1}{12cm}
The evolution of $\tilde{\zeta}_k \equiv k^{3/2} \zeta_k$ and
$\tilde{\Psi}_k \equiv k^{3/2}\Psi_k$ during inflation and reheating 
for a super-Hubble mode $k=a_0H_0$ in the case of 
$\xi=0$ with $m=10^{-6}m_{\rm pl}$ and  initial values
$\phi(0)=3m_{\rm pl}$, $\chi(0)=10^{-3}m_{\rm pl}$.
The inflationary period ends at $mt \approx 20$, after which 
the system enters the reheating stage. In this case, since the 
system is effectively identical with the single-field case, 
$\tilde{\zeta}_k$ is conserved.
{\bf Inset}: The evolution of field perturbations $\delta\tilde{\phi}_k
\equiv k^{3/2} \delta\phi_k/m_{\rm pl}$ and $\delta\tilde{\chi}_k
\equiv k^{3/2} \delta\chi_k/m_{\rm pl}$.
\end{figcaption}
\end{center}
\end{figure}

For positive $\xi$,  $\chi$ and long-wave $\delta\chi_k$
modes exponentially decrease during inflation.
This decreasing rate strongly depends on the strength of the 
coupling $\xi$. When $0<\xi<3/16$, the order $\nu$ of the 
Hankel function is in the range of $0<\nu<3/2$,
and long-wave $\delta\chi_k$ modes evolve
as $\delta\chi_k \sim a^{-(3/2-\nu)}$. In  Fig.~2,
we plot the evolution of $\delta\tilde{\chi}_k \equiv k^{3/2}
\delta\chi_k/m_{\rm pl}$ for several $\xi$ with 
initial values $\phi(0)=3m_{\rm pl}$ and $\chi(0)=10^{-3}m_{\rm pl}$.
For $\xi=0.01$, $\delta\tilde{\chi}_k$ decreases only by one order of 
magnitude during inflation. With the increase of $\xi$, the inflationary
suppression becomes relevant and the decreasing rate is getting larger.
When $\xi>3/16$ (i.e., $\nu^2<0$), super-Hubble $\delta\chi_k$ 
fluctuations decay as $\delta\chi_k \sim a^{-3/2}$ irrespective of 
the coupling $\xi$.
This means that the amplitude of $\delta\chi_k$ at the end of inflation
depends on the total amount of inflation.
In the simulation of initial values $\phi(0)=3m_{\rm pl}$ and 
$\chi(0)=10^{-3}m_{\rm pl}$, $\delta\tilde{\chi}_k$ decreases of order
$\delta\tilde{\chi}_k~\lsim~10^{-40}$ since the number of e-foldings
is about $N \sim 57$ in this case. 
The homogeneous $\chi$ field is also affected 
by this suppression. Hence $\chi$ dependent terms in 
the rhs of Eq.~$(\ref{B12})$ can be negligible relative to the $\phi$ 
dependent term, leading to the conservation of the super-Hubble 
curvature perturbation during inflation.

As is found in Eq.~$(\ref{B10})$, the scalar curvature reduces with the
decrease of inflaton, unless the $\chi$ field is amplified. 
When the kinetic term of inflaton becomes comparable 
to its potential energy, the scalar curvature begins to oscillate 
due to the oscillation of inflaton.
It is well known that $\chi$ particles coupled to $\phi$ can be 
nonperturbatively produced by parametric resonance during preheating 
era\cite{pre2}. In the present model, we can also expect the 
amplification of the $\delta\chi_k$ fluctuation due to the oscillating scalar
curvature. This geometric preheating scenario was studied in 
Ref.~\cite{nonminimalpre2,nonminimalpre3} neglecting 
metric perturbations.
In Fig.~2, we find that the $\delta\chi_k$ fluctuation undergoes 
the parametric excitation during the reheating era ($mt~\gsim~20$)
for the coupling $\xi~\gsim~10$, while there is no growth for 
$\xi~\lsim~1$. With the increase of $\xi$, the growth rate of 
$\delta\chi_k$ gets gradually larger, leading to the efficient particle
production. Nevertheless, as deeply studied in Ref.~\cite{nonminimalpre3},
larger values of $\xi$ do not necessarily result in the larger amount of 
the final fluctuation, due to the suppression effect of the $\chi$ particle
production itself.
In fact, the final variance takes the maximum value at 
$\xi=100 \sim 200$~\cite{nonminimalpre3}.
This indicates that long-wave $\delta\chi_k$ fluctuations are still
much smaller relative to inflaton fluctuations even in the case of 
$\xi~\gsim~100$ because of the very small amplitude at the end of inflation.
As a result, the existence of the preheating era does not lead to 
the enhancement of super-Hubble metric perturbations, whose source terms
in the rhs of Eq.~$(\ref{B12})$ are completely dominated by 
the inflaton-dependent term.
We have numerically checked that the curvature perturbation $\zeta_k$ and 
metric perturbations $\Psi_k$, $\Phi_k$ on large scales
exhibit the same behavior as shown in Fig.~1, 
as long as the initial value of $\chi$  at the onset of inflation 
is much smaller than $\phi$.
As a result, the adiabatic picture of large-scale cosmological perturbations 
in the single-field case still holds even during preheating in the presence of 
the positive nonminimal coupling.

\begin{figure}
\begin{center}
\singlefig{12cm}{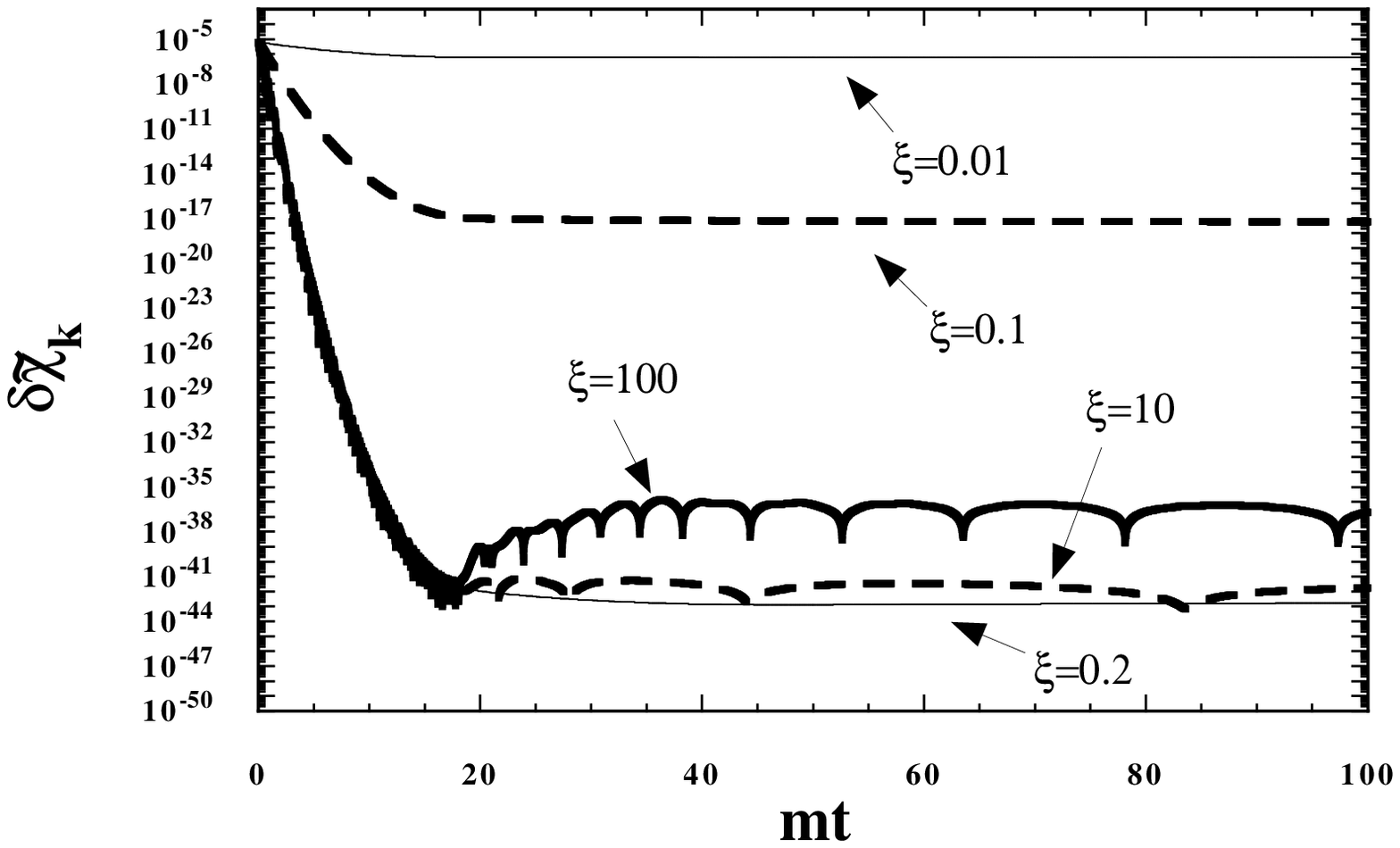}
\begin{figcaption}{Fig2}{12cm}
Suppression of the field fluctuation $\delta\tilde{\chi}_k$ 
during inflation and reheating for a 
super-Hubble mode $k=a_0H_0$ in the case of 
$\xi=0.01, 0.1, 0.2, 10, 100$ with $m=10^{-6}m_{\rm pl}$ and 
initial values $\phi(0)=3m_{\rm pl}$, $\chi(0)=10^{-3}m_{\rm pl}$. 
When $0<\xi<3/16$, $\delta\tilde{\chi}_k$
decreases as $\sim a^{(-3/2-\nu)}$, while $\delta\tilde{\chi}_k
\sim a^{-3/2}$ for $\xi>3/16$ independent of the strength of $\xi$.
The homogeneous $\chi$ field exhibits the similar behavior.
\end{figcaption}
\end{center}
\end{figure}

We should mention the evolution of $\delta\chi_k$ and $\Psi_k, \Phi_k$ 
in sub-Hubble modes during preheating.
For large $k$-modes, the adiabatic solution for 
the $\delta\chi_k$ fluctuation is estimated by Eq.~$(\ref{C8})$ as
\begin{eqnarray}
|\delta \tilde{\chi}_k| \approx \bar{k}\frac{m}{m_{\rm pl}},~~~~
|\delta \dot{\tilde{\chi}}_k|/m \approx \bar{k}^2 \frac{m}{m_{\rm pl}},
\label{C10}
\end{eqnarray}
where $\bar{k} \equiv k/(ma_I)$ with $a_I$ the scale factor 
at the beginning of preheating.
Since sub-Hubble modes correspond to $\bar{k}~\gsim~1$,
the amplitude of the $\delta\chi_k$ fluctuation at the end
of inflation is found to be
$|\delta\tilde{\chi}_k|~\gsim~m/m_{\rm pl} \sim 10^{-6}$, 
which is not strongly suppressed 
compared with the super-Hubble case.
Then the growth of the total variance 
\begin{eqnarray}
\langle \delta\chi^2 \rangle \equiv \frac{1}{2\pi^2}\int k^2 |\delta\chi_k|^2 dk
=\frac{m_{\rm pl}^2}{2\pi^2}\int |\delta\tilde{\chi}_k|^2 d (\log k), 
\label{C11}
\end{eqnarray}
is typically governed by sub-Hubble modes.
This situation is similar to the preheating scenario with the 
$\frac12 g^2\phi^2\chi^2$ coupling\cite{suppress}. 
We depict in Fig.~3 the evolution of $\delta\chi_k$, $\delta\phi_k$,
and $\Psi_k$ in preheating phase for a sub-Hubble mode
$\bar{k}=3$ with $\xi=100$.
While the $\delta\chi_k$ fluctuation exhibits parametric excitation
by the oscillating scalar curvature, we find that sub-Hubble metric
perturbations are hardly enhanced during preheating.
In spite of the unsuppressed initial conditions for sub-Hubble 
$\delta\chi_k$ and $\delta\dot{\chi}_k$ modes in the 
rhs of Eq.~$(\ref{B12})$, the homogeneous components $\chi$
and $\dot{\chi}$ are strongly damped as in the case of super-Hubble
$\delta\chi_k$ modes.
Then we can not expect the strong amplification of sub-Hubble
metric perturbations, due to the suppression of $\chi$-dependent source
terms in Eq.~$(\ref{B12})$. However, this result may change if we take 
into account second order metric perturbations~\cite{SMP} 
as in Ref.~ \cite{suppress}, which we do not consider in this paper. 
Since the rhs of Eq.~$(\ref{B13})$ can be negligible and the resonant term
is absent in the lhs, the inflaton fluctuation is also hardly amplified 
as is found in Fig.~3 even in the presence of metric perturbations.

For positive $\xi$, we conclude that the curvature perturbation in 
cosmologically relevant scales is conserved 
during inflation and preheating as long as $\chi$ is initially small
relative to $\phi$.

\begin{figure}
\begin{center}
\singlefig{12cm}{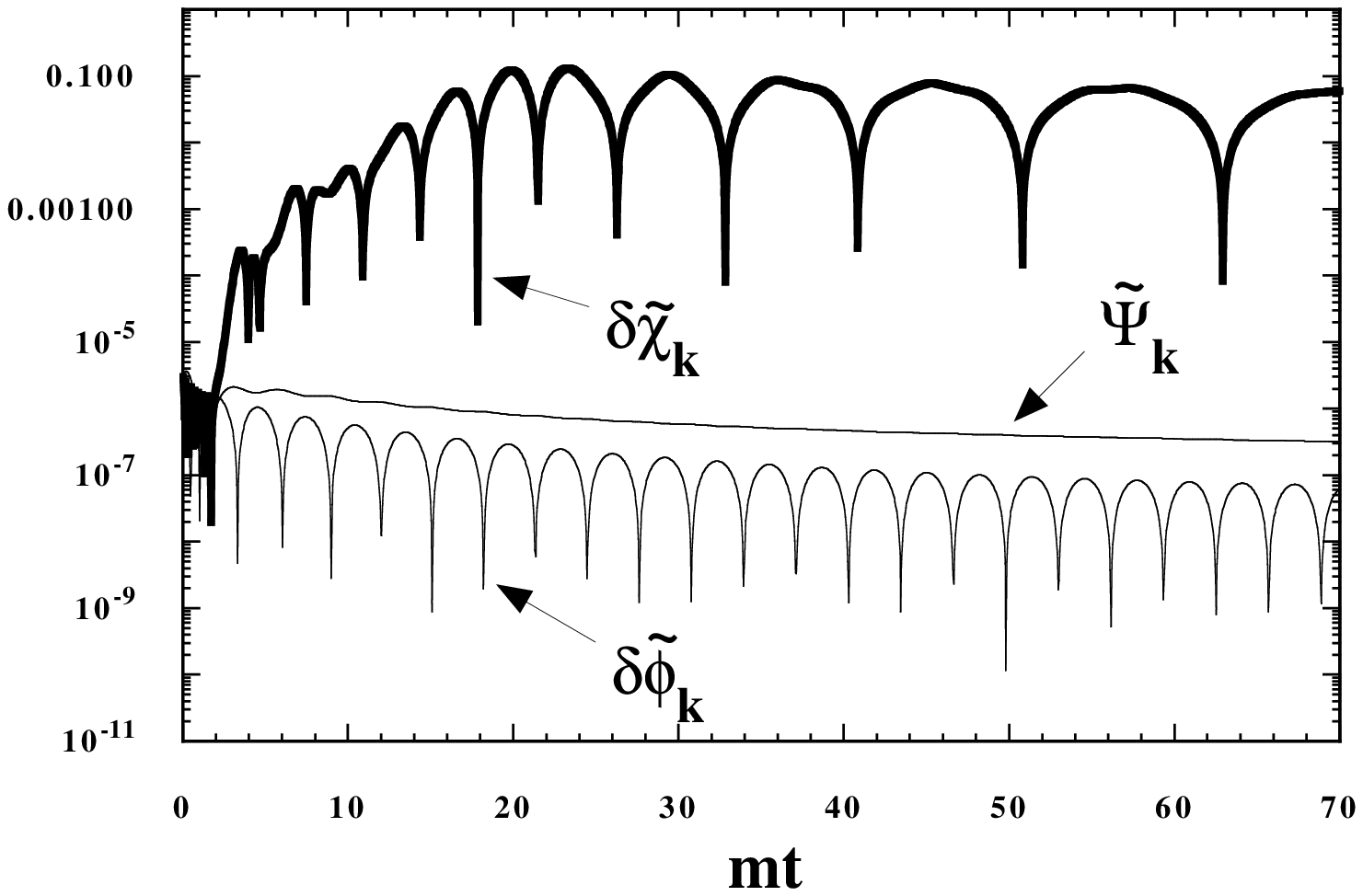}
\begin{figcaption}{Fig3}{12cm}
The evolution of $\delta\tilde{\chi}_k$, $\delta\tilde{\phi}_k$, and 
$\tilde{\Psi}_k$ during preheating for a
sub-Hubble mode $\bar{k}=3$ with $\xi=100$ and $m=10^{-6}m_{\rm pl}$.
Note that we start integrating from the end of inflation, and choose 
initial values as $\phi(0)=0.28m_{\rm pl}$ and 
$\chi(0)=10^{-40}m_{\rm pl}$ with the $\delta\chi_k$ 
fluctuation $(\ref{C10})$.
$\Phi_k$ almost coincides with $\Psi_k$ in this case. 
\end{figcaption}
\end{center}
\end{figure}

 \subsection{Case of $\xi<0$}

Let us next proceed to the case of negative $\xi$.
In this case, $\chi$ and long-wave $\delta\chi_k$
modes can be enhanced during inflation by negative instability
as is found by Eqs.~$(\ref{B9})$ and $(\ref{B14})$. 
The analytic estimation of Eq.~$(\ref{C4})$ which neglects 
metric perturbations includes the following growing solution
in small $k$-modes:
\begin{eqnarray}
|\delta\chi_k| \propto a^c,~~~~{\rm with}~~~~c=\frac32
\left(\sqrt{1+\frac{16}{3}|\xi|}-1\right).
\label{C12}
\end{eqnarray}
This growth rate gets larger with the increase of $|\xi|$.
The exponential increase of the $\delta\chi_k$ fluctuation also stimulates 
the growth of super-Hubble metric perturbations as is found in 
Eq.~$(\ref{B12})$. Then metric perturbations will strengthen 
field resonances by Eqs.~$(\ref{B13})$ and $(\ref{B14})$
in the perturbed metric case,
as we numerically study it later.
What we are mainly concerned with is how the dynamics and 
produced perturbations are modified during inflation 
and preheating by negative nonminimal coupling.

Let us first consider the case of $\phi(0)=3m_{\rm pl}$
where the number of e-foldings reaches $N \sim 57$
in the absence of the $\chi$ field.
If the negative nonminimal coupling is taken into account,
the total amount of inflation is not necessarily sufficient to 
solve cosmological puzzles.
For example, when $\chi(0)=10^{-3}m_{\rm pl}$, the dynamics of 
inflation is modified due to the enhancement of the $\chi$ field 
for $|\xi|~\gsim~0.02$.

\begin{figure}
\begin{center}
\singlefig{12cm}{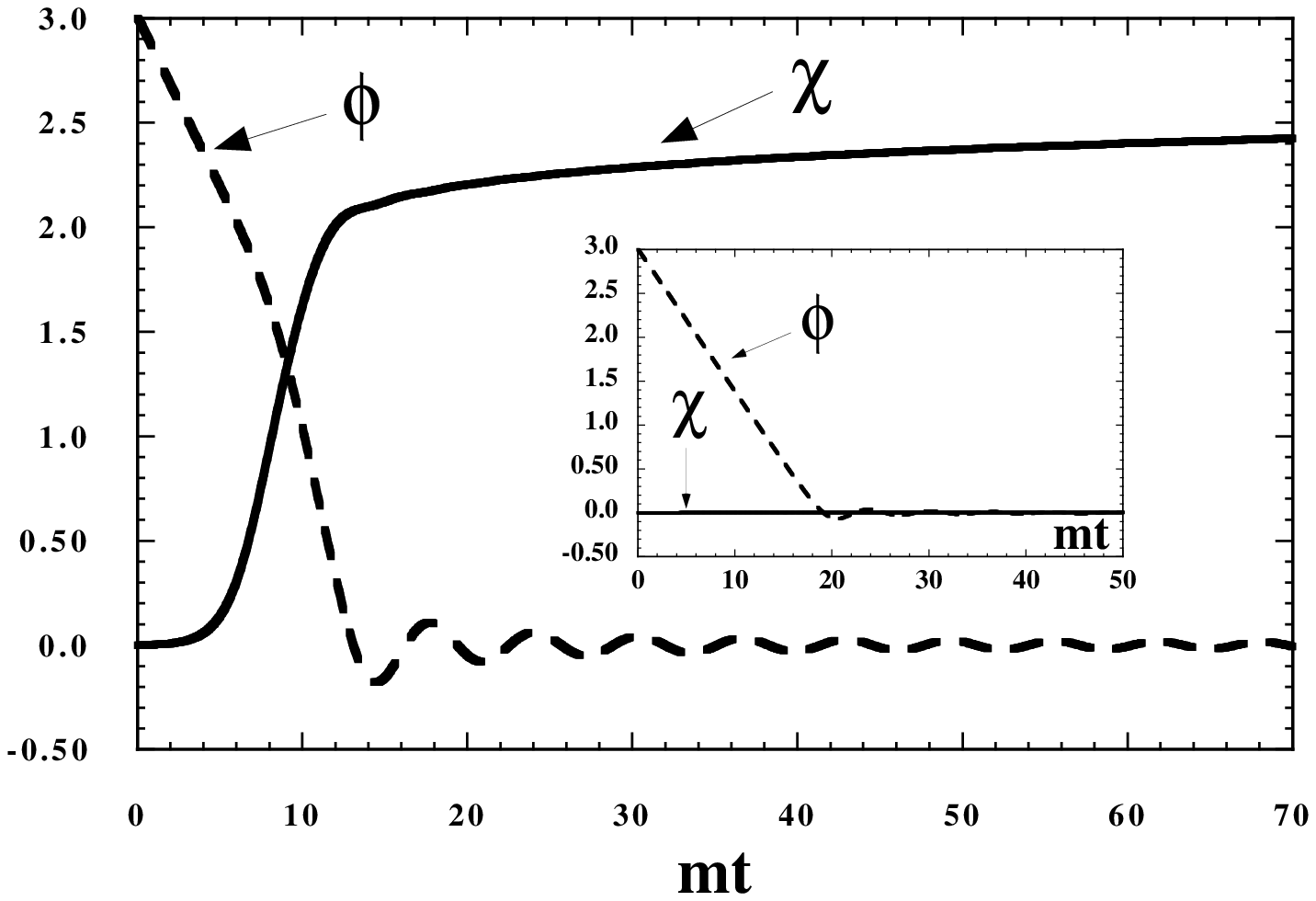}
\begin{figcaption}{Fig4}{12cm}
The evolution of $\phi$ and $\chi$ fields
during inflation and reheating  for a super-Hubble mode 
$k=a_0H_0$ in the case of $\xi=-0.05$ and initial values
$\phi(0)=3m_{\rm pl}$, $\chi(0)=10^{-3}m_{\rm pl}$
with $m=10^{-6}m_{\rm pl}$. The $\chi$ field rapidly 
grows by negative instability, whose effect terminates inflation 
at $mt \approx 13$.
{\bf Inset}: $\phi$ and $\chi$ vs $t$ for $\xi=-0.005$.
In this case, the $\chi$ field is hardly enhanced.
\end{figcaption}
\end{center}
\end{figure}

In Fig.~4, we plot the evolution of the homogeneous 
$\phi$ and $\chi$ fields for $\xi=-0.05$ with initial values
$\phi(0)=3m_{\rm pl}$ and $\chi(0)=10^{-3}m_{\rm pl}$.
The negative nonminimal coupling leads to the growth the $\chi$ field,
which catches up the $\phi$ field for $mt \approx 10$.
In the initial stage where $\phi$ is larger than $\chi$,
the dynamics of inflation is approximately described 
by Eqs.~$(\ref{C2})$ and $(\ref{C3})$.
However, after $\chi$ exceeds $\phi$ for $mt~\gsim~10$, 
these relations can no longer be applied.
With the increase of $\chi$, the potential $V(\phi)$ becomes 
gradually unimportant relative to $\chi$-dependent terms in 
Eq.~$(\ref{B7})$ and the $1/(1-\xi\kappa^2\chi^2)$ factor also 
gets smaller,  the Hubble expansion rate decreases
faster than in the $\xi=0$ case. 
Then inflation ends at $mt_e \approx 13$ with the e-folding number
$N \approx 42$. Note that these values are smaller than in the 
minimally coupled case, $mt_e \approx 20$ and $N \approx 58$.
Since the decrease of the scalar curvature $R=6(2H^2+\dot{H})$ is accompanied by 
the decrease of $H$, the growth of the $\chi$ field typically  becomes 
irrelevant after the inflationary period terminates (see Fig.~4).
We show in the inset of Fig.~4 the evolution of 
$\phi$ and $\chi$ for $\xi=-0.005$. In this case, the $\chi$ field
is hardly amplified, which results in almost the same dynamics of 
inflation as in the $\xi=0$ case. 

\begin{figure}
\begin{center}
\singlefig{12cm}{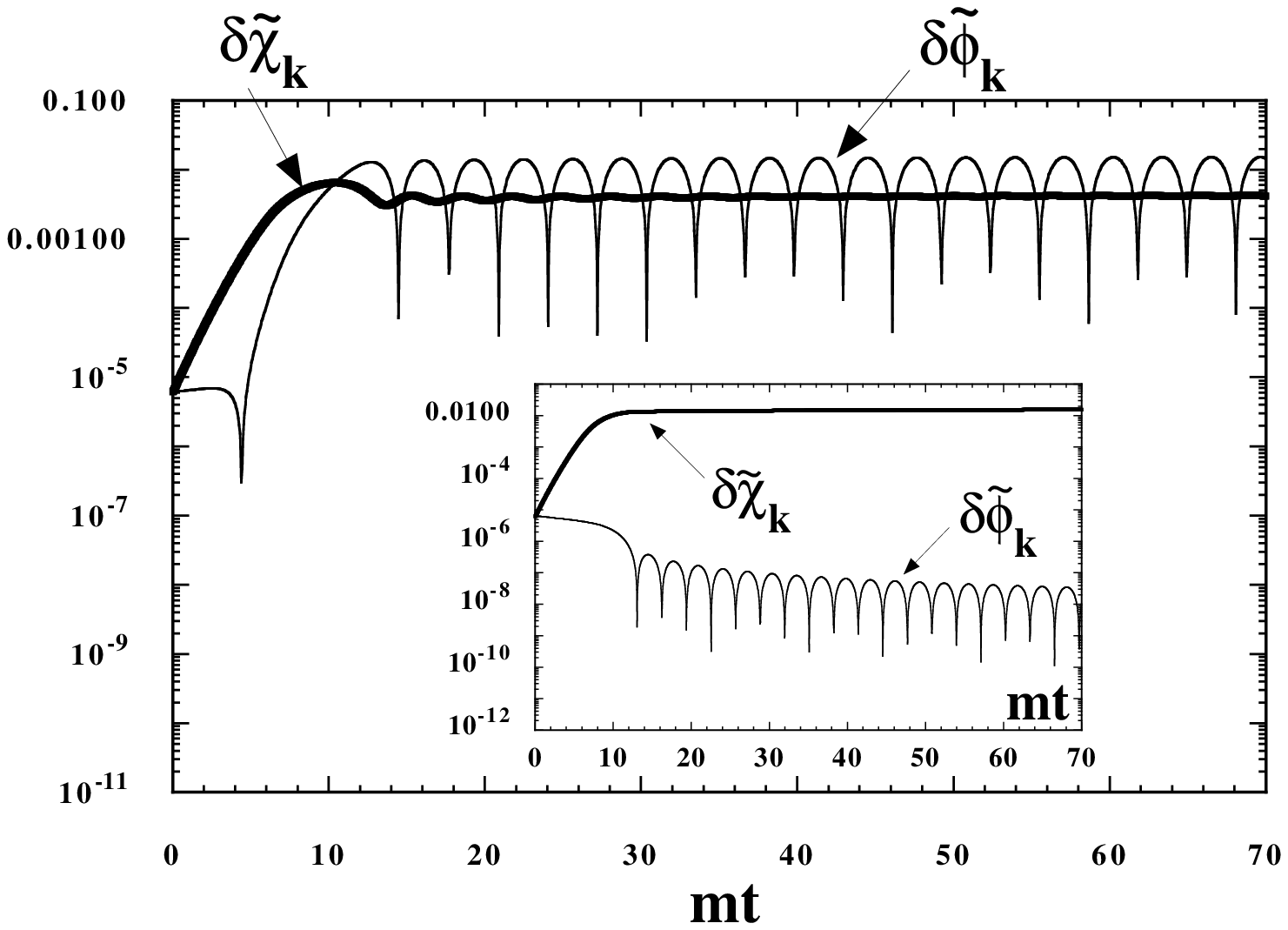}
\begin{figcaption}{Fig5}{12cm}
Growth of field fluctuations $\delta\tilde{\chi}_k$ and 
$\delta\tilde{\phi}_k$ during inflation and reheating 
for a super-Hubble mode 
$k=a_0H_0$ in the case of $\xi=-0.05$ and initial values
$\phi(0)=3m_{\rm pl}$, $\chi(0)=10^{-3}m_{\rm pl}$
with $m=10^{-6}m_{\rm pl}$. Both field fluctuations are
amplified during inflation.
{\bf Inset}: $\delta\tilde{\chi}_k$ and $\delta\tilde{\phi}_k$ 
vs $t$ for $\xi=-0.05$ neglecting metric perturbations.
Although $\delta\tilde{\chi}_k$  is enhanced as in the perturbed 
metric case, $\delta\tilde{\phi}_k$ does not grow in the absence
of metric perturbations.
\end{figcaption}
\end{center}
\end{figure}

In Fig.~5 we show the evolution of large-scale $\delta\chi_k$ 
and $\delta\phi_k$ fluctuations for $\xi=-0.05$ with
$\phi(0)=3m_{\rm pl}$ and $\chi(0)=10^{-3}m_{\rm pl}$.
We numerically found that the growth of field perturbations is relevant
for $|\xi|~\gsim~0.02$.
In the case of $\xi=-0.05$, $\delta\chi_k$ fluctuations in small $k$-modes
are enhanced from the beginning as described in Eq.~$(\ref{C12})$ with 
$c\approx 0.188$. Recalling that the total amount of inflation is about 
$N \approx 42$, $|\delta\chi_k|$ is amplified about $10^3$ 
times during inflation by  the analytic estimation of Eq.~$(\ref{C12})$  
neglecting metric perturbations. 
We plot in the inset of Fig.~5 the evolution of long-wave 
field perturbations in the unperturbed metric case [i.e., setting $\Psi_k=
\Phi_k=0$ in Eqs.~$(\ref{B13})$ and $(\ref{B14})$].
We can easily confirm that numerical calculations coincide
with the analytic result $(\ref{C12})$ fairly well.
In the perturbed metric case, 
the evolution of the $\delta\chi_k$ fluctuation is almost the same as in the
unperturbed metric case except for the final short stage of inflation,
which indicates that the $\xi R$ term in the lhs of Eq.~$(\ref{B14})$ 
mainly determines the growth of $\delta\chi_k$ even taking into account 
metric perturbations.
 
In contrast, the difference appears in the inflaton fluctuation.
If we neglect metric perturbations, the $\delta\phi_k$ fluctuation 
does not grow nonperturbatively in the massive inflaton model.
Including metric perturbations, the enhancement of $\chi$ and 
$\delta\chi_k$ fluctuations in small $k$-modes stimulates the growth 
of super-Hubble metric perturbations by Eq.~$(\ref{B12})$.
Then the rhs of Eq.~$(\ref{B13})$ leads to the amplification of the 
$\delta\phi_k$ fluctuation, which is absent in the rigid spacetime
case. This difference is clearly seen in Fig.~5.

\begin{figure}
\begin{center}
\singlefig{12cm}{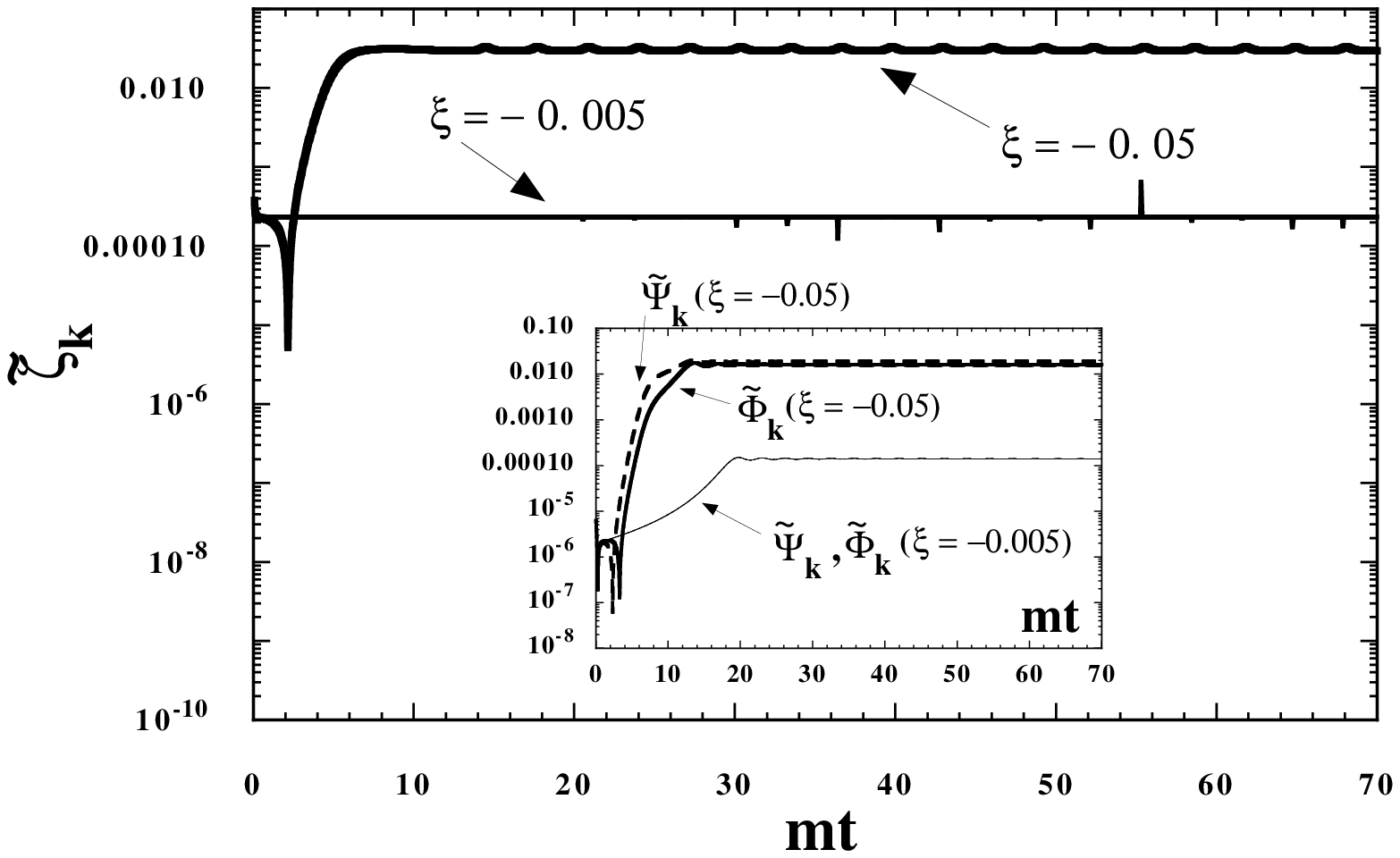}
\begin{figcaption}{Fig6}{12cm}
The evolution of the curvature perturbation $\tilde{\zeta}_k$
during inflation and reheating for a super-Hubble mode 
$k=a_0H_0$ in the cases of $\xi=-0.05$ and $\xi=-0.005$ with 
$\phi(0)=3m_{\rm pl}$, $\chi(0)=10^{-3}m_{\rm pl}$, and
$m=10^{-6}m_{\rm pl}$. We find that $\tilde{\zeta}_k$ 
grows nonperturbatively during inflation for $\xi=-0.05$, 
while it is conserved for $\xi=-0.005$. 
{\bf Inset}: The evolution of super-Hubble metric perturbations 
$\tilde{\Psi}_k$ and $\tilde{\Phi}_k$ 
for $\xi=-0.05$ and $\xi=-0.005$. 
\end{figcaption}
\end{center}
\end{figure}

Let us investigate the evolution of the curvature perturbation 
$\zeta_k$ and  metric perturbations $\Psi_k$ and $\Phi_k$
on large scales.
Since the growth of metric perturbations is accompanied by 
the excitement of the $\chi$ field fluctuation, $\zeta_k$ increases 
during inflation when $|\xi|~\gsim~0.02$ with initial 
values $\phi(0)=3m_{\rm pl}$ and $\chi(0)=10^{-3}m_{\rm pl}$.
In Fig.~6, we show the evolution of $\Psi_k$ and $\Phi_k$
for $\xi=-0.05$ and $\xi=-0.005$.
When $\xi=-0.005$, $\Psi_k$ almost coincides
with $\Phi_k$  as in the case of $\xi \ge 0$.
For $\xi=-0.05$, however, super-Hubble metric perturbations 
are strongly amplified during inflation, leading to the distortions
in the CMB spectrum. Note that the difference of 
$\Psi_k$ and $\Phi_k$ appears in this case, due to the enhancement
of the $\chi$ field fluctuation.
While the super-Hubble curvature perturbation is conserved
for $\xi=-0.005$, it exhibits rapid growth during inflation 
for $\xi=-0.05$.
This means that the standard picture of adiabatic perturbations in the  
single-field case can no longer be applied in the presence of the 
nonminimally coupled $\chi$ field with negative coupling.

In TABLE I, we show the number of e-foldings, the homogeneous
$\chi$ field, and the super-Hubble curvature perturbation 
at the end of inflation for several values of $\xi$ with initial 
conditions $\phi(0)=3m_{\rm pl}$ and $\chi(0)=10^{-3}m_{\rm pl}$. 
Since the enhancement of the $\chi$ field is weak for 
$|\xi|~\lsim~0.02$, $\zeta_k$ remains constant during inflation.
For $|\xi|~\gsim~0.02$, the rapid growth of $\chi$ makes
the inflationary period terminate earlier as confirmed in TABLE I.
This leads to the smaller amount of inflation with the increase 
of $|\xi|$. For example, when $\xi=-1$, the system soon enters 
the reheating stage after only 4 e-foldings.
We have to caution that large $|\xi|$ does not necessarily
yield the larger values of $\chi$ and $\zeta_k$ at the end of inflation,
because the duration of inflation gets shorter.
In fact, $\zeta_k$ decreases with the increase of $|\xi|$ for 
$|\xi|~\gsim~0.05$, although large $|\xi|$ also leads to the 
distortions in the anisotropies of CMB.
The important point is that the successful inflationary scenario
can be completely violated with the existence of the large negative 
nonminimal coupling.

\begin{figure}
\begin{center}
\singlefig{12cm}{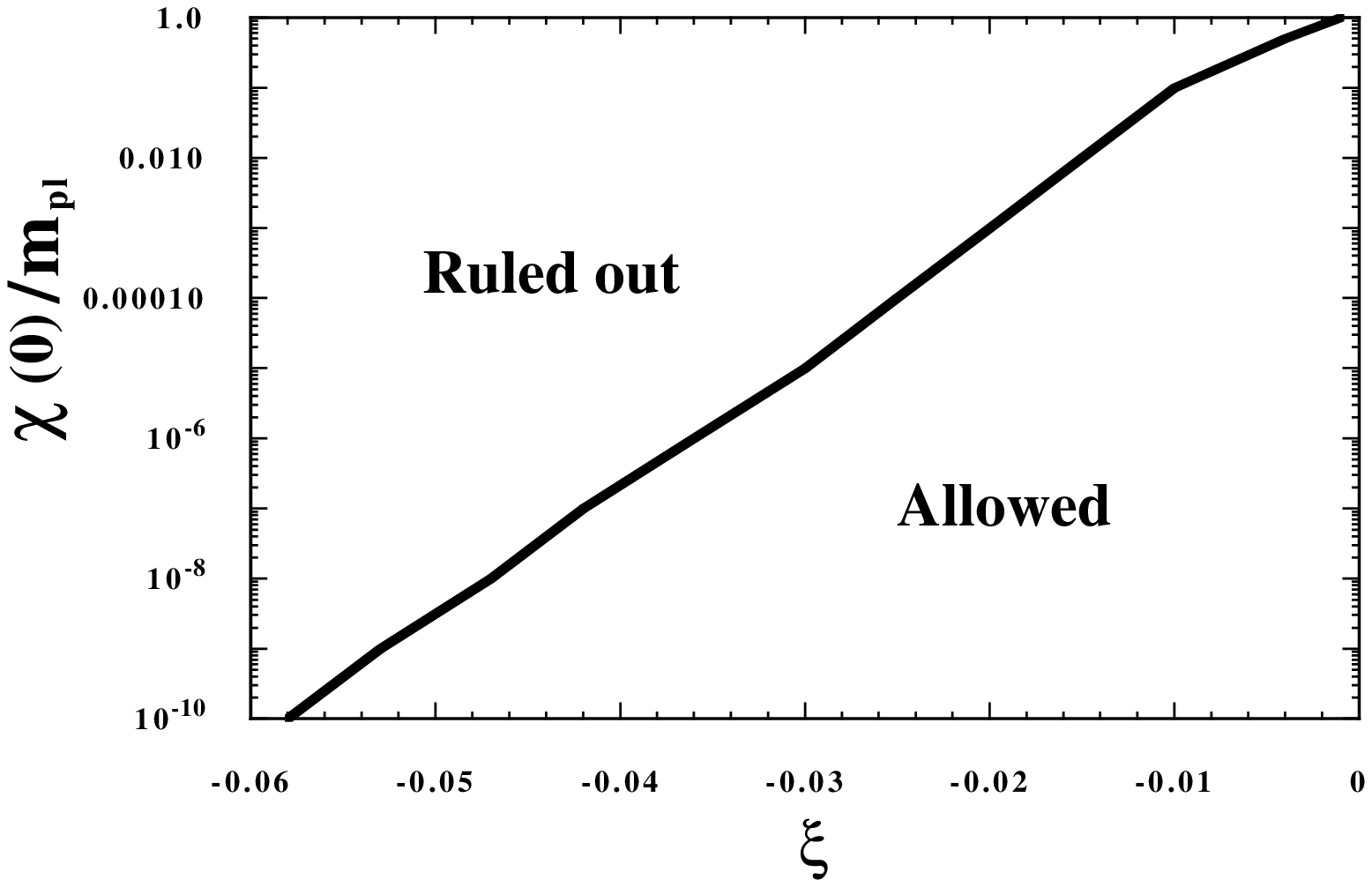}
\begin{figcaption}{Fig7}{12cm}
The parameter regions of the coupling $\xi$ and the initial 
$\chi(0)$ where the inflationary scenario proceeds in successful way 
or not, for the case of $\phi(0)=3m_{\rm pl}$ and 
$m=10^{-6}m_{\rm pl}$. We find that large negative coupling 
prevents successful inflation unless we choose small values of $\chi(0)$.  
\end{figcaption}
\end{center}
\end{figure}

Let us consider the case where initial values of $\chi$ are changed.
With the decrease of $\chi(0)$, larger $|\xi|$
is required for the growth of curvature perturbations.
When $\chi(0)$ is close to the order of $m_{\rm pl}$,
$\chi$ soon catches up inflaton even for not so large values of 
$|\xi|$, which prevents successful inflation. In this case, inflation 
typically terminates with small amount of e-foldings before $\zeta_k$ 
begins to grow significantly.
We present two-dimensional plots of $\xi$ and
$\chi(0)$ which divide the ``allowed" and ``ruled out " regions 
in Fig.~7 for the case of $\phi(0)=3m_{\rm pl}$. 
The ``allowed" regions mean that conditions
of $\delta_H(k)<2\times 10^{-5}$ and $N>55$ are both satisfied 
at the end of inflation for the inflaton mass $m=10^{-6}m_{\rm pl}$.
When $\chi(0)~\lsim~0.1m_{\rm pl}$, the separating curve is mainly
determined by the condition of $\delta_H(k)<2\times 10^{-5}$
(i.e., $\zeta_k$ remains almost constant on super-Hubble scales).
For $\chi(0)~\gsim~0.1m_{\rm pl}$, the condition of $N>55$ plays 
the dominant role rather than that  of the density perturbation.
It is important to note that wide ranges of parameters are ruled out 
even in the case of $|\xi|~\lsim~0.1$ unless we take smaller values 
of $\chi(0)$. When $|\xi|~\gsim~1$, we find that nonlinear growth of 
super-Hubble curvature perturbations is inevitable even
for very small initial $\chi$ as $\chi(0)=10^{-50}m_{\rm pl}$.

One may consider that allowed regions may become wider if initial 
values of inflaton are  larger. However, this is not generally true. 
Since larger values of $\phi$ correspond to the larger potential energy,  
the inflationary period during which the $\phi$ field dominates the 
dynamics of the system is longer. This prolonged inflation leads to
the amplification of super-Hubble $\zeta_k$ as well as
the $\chi$ field fluctuation.
For example, when $\phi(0)=4m_{\rm pl}$ and $\chi(0)=10^{-3}m_{\rm pl}$
with $\xi=-0.02$, $\zeta_k$ grows up to $\tilde{\zeta}_k \approx 0.07$,
while for the smaller value $\phi(0)=3m_{\rm pl}$, $\zeta_k$ remains 
almost constant for the same value of $\xi$. 
For $\phi(0)=4m_{\rm pl}$ and $\chi(0)=10^{-3}m_{\rm pl}$, 
the allowed values of $\xi$ are found to be $|\xi|~\lsim~0.01$,
whose condition is tighter than in the case of $\phi(0)=3m_{\rm pl}$.
When $\chi(0)$ is close to of order unity, 
larger values of $\phi(0)$ typically make the e-folding number larger,
which can broaden allowed regions in some cases.  
For example, when $\phi(0)=4m_{\rm pl}$ and $\chi(0)=m_{\rm pl}$, 
the allowed values are $|\xi|~\lsim~0.007$, while $|\xi|~\lsim~0.001$
for $\phi(0)=3m_{\rm pl}$ and $\chi(0)=m_{\rm pl}$.
However, when $\phi(0)$ takes further large values as 
$\phi(0)~\gsim~10m_{\rm pl}$, $\zeta_k$ grows nonadiabatically
even for $\xi=-0.001$.
This indicates that large $\phi(0)$ does not necessarily result 
in the successful inflation in the presence of negative
nonminimal coupling.

In the reheating phase, parametric amplification of the $\chi$
fluctuation is relevant only for the case of 
$\xi~\lsim~-1$\cite{nonminimalpre2,nonminimalpre3},
because the scalar curvature gradually decreases during inflation.
However, such large values of $|\xi|$ generally prevents
the successful inflationary scenario as explained above,
which will be ruled out even if $\chi(0)$ is initially very small.
This is in contrast with the positive $\xi$ case where
exponential suppression of long-wave modes in the $\delta\chi_k$ 
fluctuation do not affect on the dynamics of inflation.
In the absence of other interactions, negative nonminimal coupling leads 
to the strong distortions on CMB in wide ranges of parameters.

 \subsection{The $\frac12g^2\phi^2\chi^2$ coupling is taken into account}

So far, we have not considered the interaction between $\phi$ and 
$\chi$ fields. Taking into account the simple four-point coupling 
$\frac12 g^2\phi^2\chi^2$ provides a way to escape nonadiabatic 
growth of super-Hubble curvature perturbations.
The background equations for the scale factor and inflaton 
are obtained by changing $V(\phi)=\frac12 m^2\phi^2$ to
$V(\phi,\chi)=\frac12 m^2\phi^2+\frac12 g^2\phi^2\chi^2$ in
Eqs.~$(\ref{B7})$ and $(\ref{B8})$.
The homogeneous $\chi$ and the $\delta\chi_k$
fluctuation satisfy
\begin{eqnarray}
\ddot{\chi}+3H\dot{\chi}+(g^2\phi^2+\xi R)\chi =0,
\label{D2}
\end{eqnarray}
\begin{eqnarray}
\delta\ddot{\chi}_k+3H\delta\dot{\chi}_k+
\left( \frac{k^2}{a^2} +g^2\phi^2+\xi R \right)\delta \chi_k=
2(\ddot{\chi}+3H \dot{\chi})\Phi_k+\dot{\chi}
(\dot{\Phi}_k+3\dot{\Psi}_k)
-\xi\chi \delta R_k -2g^2\phi\chi\delta\phi_k.
\label{D3}
\end{eqnarray}
Then the effective mass of the $\chi$ and super-Hubble $\delta\chi_k$
modes are given by
\begin{eqnarray}
m_{\rm eff}^2=g^2\phi^2+\xi R.
\label{D4}
\end{eqnarray}
Neglecting the contribution of the $\chi$ field relative to inflaton
in Eq.~$(\ref{B10})$, the scalar curvature is approximately written as
$R \approx 2\kappa^2m^2\phi^2$ during inflation, which yields
the relation
\begin{eqnarray}
m_{\rm eff}^2 \approx \left[g^2+16\pi\xi
\left(\frac{m}{m_{\rm pl}}\right)^2\right]\phi^2.
\label{D5}
\end{eqnarray}
When $\xi<0$, the negative effective mass leads to the exponential 
increase of $\chi$ and long-wave $\delta\chi_k$ modes during inflation. 
This effect is weakened  in the presence of
the $g$ coupling. Especially for the positive effective mass, 
which corresponds to   
\begin{eqnarray}
g>4\sqrt{\pi} \frac{m}{m_{\rm pl}} \sqrt{|\xi|},
\label{D6}
\end{eqnarray}
the $\chi$ particle production is shut off in inflationary phase.

Neglecting metric perturbations, we have the analytic solution 
$(\ref{C4})$, where the order of the Hankel functions
is given by\cite{SH}
\begin{eqnarray}
\nu^2=\frac94-12\xi-\frac{g^2\phi^2}{H^2}.
\label{D7}
\end{eqnarray}
Since the Hubble expansion rate is approximately written as
$H^2\approx 4\pi/3(m/m_{\rm pl})^2\phi^2$ during inflation,
we obtain
\begin{eqnarray}
\nu^2 \approx \frac94-12\xi-\frac{3g^2}{4\pi}
\left( \frac{m_{\rm pl}}{m}\right)^2 .
\label{D8}
\end{eqnarray}
Eq.~$(\ref{D6})$ corresponds to cancelling the 
second term in Eq.~$(\ref{D8})$ by the $g$ term.
When $\nu^2<0$, i.e.,
\begin{eqnarray}
g>4\sqrt{\pi} \frac{m}{m_{\rm pl}} \sqrt{|\xi|+\frac{3}{16}},
\label{D9}
\end{eqnarray}
$\chi$ modes exponentially decrease as $\propto a^{-3/2}$
in the similar way as the large positive $\xi$ case.
When $g$ ranges in the region of 
$4\sqrt{\pi}(m/m_{\rm pl}) \sqrt{|\xi|}<g<
4\sqrt{\pi}(m/m_{\rm pl})\sqrt{|\xi|+3/16}$, 
$\chi$ decays more slowly as $\propto a^{-(3/2-\nu)}$.
However, as long as the condition $(\ref{D6})$ is satisfied and 
$\chi$ is initially small relative to $\phi$, the $\chi$ field hardly 
affects the dynamics of inflation, 
which results in the conservation of the curvature perturbation
$\zeta_k$ on super-Hubble scales. 

Let us consider concrete cases. In Fig.~8 we plot the evolution 
of a long-wave $\delta\chi_k$ mode for $\xi=-0.05$ and several 
values of $g$ with initial conditions $\phi(0)=3m_{\rm pl}$
and $\chi(0)=10^{-3}m_{\rm pl}$.
When $g=0$, the $\delta\chi_k$ fluctuation exhibits exponential increase 
during inflation, leading to the nonadiabatic growth of super-Hubble 
curvature perturbations.
In the presence of the $g$ coupling, the conditions of Eqs.~$(\ref{D6})$ 
and $(\ref{D9})$ yield $g>1.6 \times 10^{-6}$ 
and $g>3.5 \times 10^{-6}$, respectively, 
for the typical mass scale $m=10^{-6}m_{\rm pl}$.
As is found in Fig.~8, $\delta\chi_k$ decreases very rapidly as
$\propto a^{-3/2}$ for $g=5.0 \times 10^{-6}$, while its decreasing rate is
smaller for $g=2.0 \times 10^{-6}$.
In both cases, however, large-scale curvature perturbations 
remain almost constant during inflation and reheating
(see the inset of Fig.~8).

\begin{figure}
\begin{center}
\singlefig{12cm}{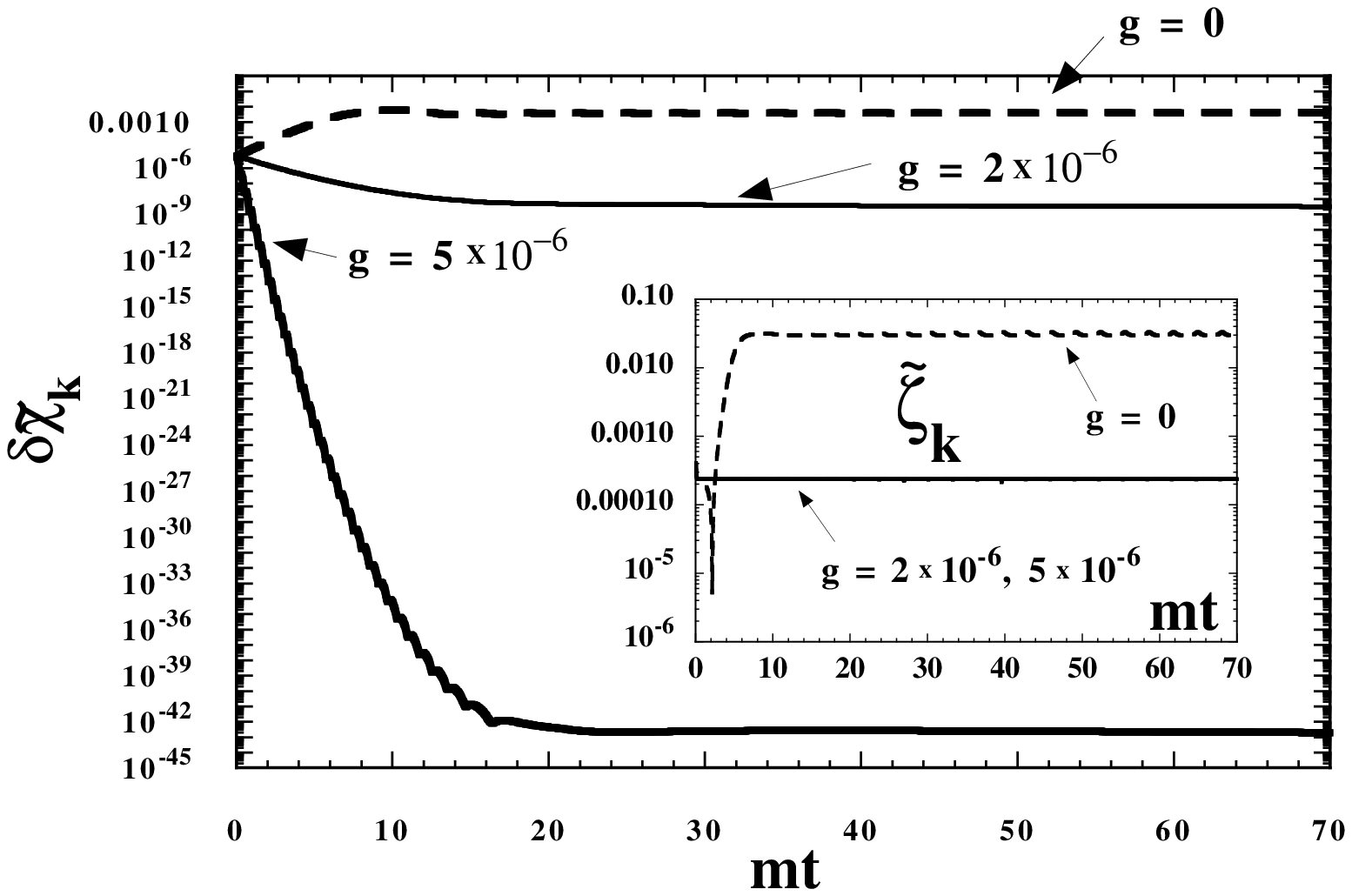}
\begin{figcaption}{Fig8}{12cm}
The evolution of the field fluctuation $\delta\tilde{\chi}_k$
during inflation and reheating for a 
super-Hubble mode $k=a_0H_0$ in the cases of $g=0$, $2 \times10^{-6}$,
and $5 \times 10^{-6}$ with $\xi=-0.05$, $m=10^{-6}m_{\rm pl}$, and 
initial values $\phi(0)=3m_{\rm pl}$, $\chi(0)=10^{-3}m_{\rm pl}$. 
{\bf Inset}: $\tilde{\zeta}_k$ vs $t$ for $g=0$, $2 \times10^{-6}$,
and $5 \times 10^{-6}$. For the values of $g$ which satisfy the relation 
$(\ref{D6})$, i.e, $g>1.6 \times 10^{-6}$, the curvature perturbation
is conserved in large scales. 
\end{figcaption}
\end{center}
\end{figure}

In the case of $|\xi| \gg 1$, Eq.~$(\ref{D9})$ approximately 
takes the form of Eq.~$(\ref{D6})$, which reads 
$g~\gsim~7.1 \times 10^{-6}\sqrt{|\xi|}$ for the inflaton mass 
$m=10^{-6}m_{\rm pl}$. Even for very large values of $|\xi|$ such as 
$\xi=-10^4$, the $\xi$ effect can be removed for 
$g~\gsim~7.1 \times 10^{-4}$. 
If the coupling between $\phi$ and $\chi$ is greater than of order 
$10^{-3}$, the $\chi$ particle production 
during inflation discussed in Ref.~\cite{SH} will be irrelevant.
When $g~\gsim~10^{-3}$, since the $g$ effect is typically dominant
relative to the negative nonminimal coupling unless $|\xi|$ is 
unnaturally large, $\chi$ particles are created
during preheating in the usual manner due to the 
$g$ resonance\cite{pre2}. 
Since long-wave $\chi$ modes are
exponentially suppressed during inflation for the case of 
$g \gg 7.1 \times 10^{-6}\sqrt{|\xi|}$,  the existence of
the preheating era does not lead to the amplification of super-Hubble
metric perturbations, which provides the standard conservation law of 
large-scale curvature perturbations.

 \section{Conclusions}

We have studied the dynamics and perturbations in the multi-field 
inflation with a nonminimally coupled $\chi$ field.
When the coupling $\xi$ is positive, $\chi$
and long-wave $\delta\chi_k$ fluctuations are exponentially 
suppressed in de Sitter background. 
In this case, the existence of the $\chi$ field hardly affects the 
dynamics of inflation, and  the ordinary adiabatic 
scenario in large-scale curvature perturbations is not modified
as long as $\chi$ is initially small relative to inflaton.
Although $\chi$ fluctuations grow by parametric resonance
during preheating after inflation 
for large values of $\xi (\gg 1)$, this process is not sufficient 
to enhance super-Hubble metric perturbations, since the 
inflationary suppression is strong. 

In contrast, negative nonminimal coupling can lead to the strong 
inflationary $\chi$ particle production in long-wave modes.
This exponential increase of the $\chi$ fluctuation makes 
super-Hubble metric perturbations grow too, 
which violates the standard conservation property of 
large-scale curvature perturbations in adiabatic inflation models.
We find that even the coupling $|\xi|$ less than unity yields
the exponential growth of the $\chi$ fluctuation in small $k$-modes,
which terminates the inflationary period earlier than 
in the minimally coupled case.
This effect reduces the total amount of inflation (e-foldings),
in addition to the nonadiabatic increase of super-Hubble 
curvature perturbations.
Large values of $|\xi|$ greater than unity make the inflationary phase
very short, whose amount of inflation is typically insufficient 
for the success of the inflationary scenario.  
We have constrained the strength of negative nonminimal coupling by two 
requirements that the large-scale curvature perturbation is almost conserved 
and the number of e-foldings satisfies $N>55$.
Since the evolution of the $\chi$ fluctuation depends on its initial value 
at the beginning of inflation, we examined the allowed regions in 
two-dimensional plots of $\xi$ and $\chi(0)$.
With the increase of $|\xi|$, we require smaller values of $\chi(0)$
for the successful inflationary scenario. When $|\xi|~\gsim~1$,
we find that strong enhancement of large-scale curvature perturbations
is inevitable even for very small values of $\chi(0)$. 

As one escape route from nonadiabatic growth of super-Hubble
metric perturbations, we considered the interaction 
$\frac12g^2\phi^2\chi^2$ between $\phi$ and $\chi$ fields.
Introducing this coupling makes the effective $\chi$ mass heavy,
which suppresses the inflationary $\chi$ particle production 
by negative nonminimal coupling. 
If two couplings satisfy the condition $(\ref{D6})$, 
the $\chi$ fluctuation does not exhibit exponential increase during inflation.
This protects super-Hubble curvature perturbations from being amplified, 
because the system is effectively dominated by inflaton in this case.

Although we have considered the simple massive inflationary 
model, strong enhancement of 
long-wave $\chi$ fluctuations by negative nonminimal coupling 
will occur in potential-independent way in de Sitter background. 
Since the scalar curvature is approximately 
written as $R\approx 4\kappa^2V(\phi)$ during inflation, 
super-Hubble curvature perturbations as well as $\chi$ fluctuations 
will also grow nonperturbatively in other models of inflation  
while the potential energy $V(\phi)$ slowly decreases.
In addition to this, exponential increase of cosmological perturbations
leads to the production of inflaton particles.
In  spinodal inflation models\cite{spinodal} where the second 
derivative of $V(\phi)$ changes sign, 
the inflaton fluctuation in small $k$-modes exhibits exponential increase
when $V''(\phi)<0$ even in the single-field case. 
It is of interest to study the dynamics and perturbations in this model
with a nonminimally coupled $\chi$ field, in which amplifications
of the inflaton fluctuation may further strengthen super-Hubble metric
perturbations. 

There are other issues which we did not address in this paper.
Since negative nonminimal coupling works to violate the scale-invariance
of the CMB spectrum, we should also consider the spectral index to
constrain the strength of $\xi$ by observations.
 In the single-field case, the spectral tilts are evaluated in 
generalized Einstein theories in Ref.~\cite{spectral}, which may be 
interesting to extend to the multi-field case including the nonminimally 
coupled $\chi$ field.
In addition to this, although we did not consider backreaction effects
in this paper, detailed studies including second order
metric perturbations\cite{SMP} will be needed toward complete
understanding of the multi-field inflation and preheating.
These issues are left to future works.

\section*{ACKOWLEDGMENTS}
We thank Bruce A. Bassett for detailed and insightful comments
and Eiichiro Komatsu, Kei-ichi Maeda, and Takashi Torii
for useful discussions. We also thank David I. Kaiser for 
providing us a useful note\cite{Kaiser}. This work was supported partially 
by a Grant-in-Aid for  Scientific Research Fund of the Ministry of Education, 
Science and Culture (No. 09410217), and by the Waseda University 
Grant for Special Research Projects.

\vspace{1cm}
\begin{table}
\caption{The time $mt_f$, the number of e-foldings $N$,
the value $\chi_f$, and the super-Hubble curvature perturbation
$\tilde{\zeta}_k$ at the end of inflation with initial values
$\phi(0)=3m_{\rm pl}$ and $\chi(0)=10^{-3}m_{\rm pl}$.
With the increase of $|\xi|$, the duration
of inflation becomes shorter, which results in the smaller 
amount of e-foldings. We also find that large values of 
$|\xi|$ ($|\xi|>0.02$) leads to the nonadiabatic 
increase of the curvature perturbation due to 
the enhancement of the $\chi$ field.
}

\vskip .3cm
\noindent
\begin{tabular}{crcccccclc}
 ~& $\xi$ & ~& $mt_f$ &~& $N$
         & $\chi_f/m_{\rm pl}$ &~& $\tilde{\zeta}_k$ &~\\
        \hline
 ~&$-0.005$ & ~& 18 & ~& 57 & 0.0031 &~ & 0.00023 &~\\
 ~& $-0.01$ & ~& 18 & ~& 57 & 0.0094  &~ & 0.00023 &~\\  
 ~& $-0.05$ & ~& 12  & ~& 41& 2.0 &~ & 0.031 &~\\   
 ~& $-0.1$ & ~& 8  & ~& 23 & 2.7 &~ & 0.018 &~\\   
 ~& $-1$ & ~& 3 &~& 4 & 1.7 &~ & 0.0028 &~
\end{tabular}
\end{table}

\end{document}